\documentclass[a4paper,11pt]{article}
\pdfoutput=1 

\usepackage{jheppub} 

\usepackage{amsmath,epsf,amssymb,mathtools,latexsym,amsthm,setspace,array,pifont,hyperref,amsfonts,dsfont,cancel,braket,graphicx,booktabs,tensor,diagbox,makecell,lipsum}

\newcommand{\beq}{\begin{equation}}
\newcommand{\eeq}[1]{\label{#1}\end{equation}}
\newcommand{\bea}{\begin{eqnarray}}
\newcommand{\eea}[1]{\label{#1}\end{eqnarray}}
\renewcommand{\Im}{{\rm Im}\,}
\renewcommand{\Re}{{\rm Re}\,}

\def\mO{\mathcal{O}}

\def\zb{\bar{z}}

\def\mb{\bar{m}}

\def\pa{\partial}

\renewcommand{\[}{\begin{equation}\begin{aligned}}
\renewcommand{\]}{\end{aligned}\end{equation}}

\def\g5{\gamma_5}

\def\sigmab{\bar{\sigma}}

\def\Mt{\tilde{M}}

\def\Tt{\tilde{T}}

\def\Ct{\tilde{C}}

\def\mbt{\tilde{m}_B}

\def\b[#1]{\bold{#1}}
\def\bb[#1]{\overline{\bold{#1}}}
\def\bs[#1,#2]{\bold{#1}_{#2}}
\def\bbs[#1,#2]{\overline{\bold{#1}}_{#2}}

\def\s2{\sigma_2}
\def\ep{\epsilon}

\def\gammaflat{ \gamma_{z\zb}}
\def\gammaflatt{ \gamma^{z\zb}}

\def\paz{\pa_z}

\def\pazb{\pa_{\zb}}

\def\ketd[#1]{\ket{#1}_{\text{dressed}}}
\def\brad[#1]{\bra{#1}_{\text{dressed}}}
\def\ketas[#1]{\ket{#1}_{\text{Asymptotic}}}
\def\braas[#1]{\bra{#1}_{\text{Asymptotic}}}

\def\sigmaz{\sigma^{0}}
\def\sigmazb{\bar{\sigma}^{0}}
\def\ethb{\bar{\eth}}

\def\gammaflat{ \gamma_{z\zb}}

\def\paz{\pa_z}

\def\pazb{\pa_{\zb}}

\def\eq{\begin{equation}}
\def\eqe{\end{equation}}
\def\eqa{\begin{eqnarray}}
\def\eqae{\end{eqnarray}}

\def\gc[#1,#2,#3]{\tensor{\Gamma}{_{#1#2}^{#3}}}
\def\torsion[#1,#2,#3]{\tensor{S}{_{#1#2}^{#3}}}
\def\contorsion[#1,#2,#3]{\tensor{K}{_{#1#2}^{#3}}}
\def\chris[#1,#2,#3]{\left\{\begin{array}{c}#1 \\#2#3 \end{array}\right\}}

\title{\boldmath Subleading BMS Charges and The Lorentz Group}
\author{Uri Kol}
\affiliation{Center for Cosmology and Particle Physics, Department of Physics, New York University, 726 Broadway, New York, NY 10003, USA}
\emailAdd{urikol@gmail.com}

\abstract{
The extended BMS group includes supertranslation, dual supertranslation and Lorentz transformations.
The generators of these symmetries can be classified according to their parity into "electric" and "magnetic" types.
Using a multipole expansion of gravitational sources in the Newman-Penrose formalism, we associate each one of these charges with a particular moment.
At leading order, the "electric" and "magnetic" monopole moments correspond respectively to supertranslations and dual supertranslations.
At the first subleading order, the "electric" and "magnetic" dipole moments correspond respectively to boosts and rotations, therefore comprising the entire Lorentz group.
Electric-magnetic type of duality then rotates the boost and rotation generators into each other.
}

\begin{document} 
\maketitle
\flushbottom

\section{Introduction}\label{sec:intro}

The isometry group of asymptotically flat spacetimes, known as the Bondi-Metzner-Sachs (BMS) group, includes Lorentz transformations and supertranslations \cite{Bondi:1960jsa,Bondi:1962px,Sachs:1962wk}.
Recently, a new \emph{dual supertranslation} symmetry of asymptotically flat spacetimes was discovered in \cite{Kol:2019nkc}.
The corresponding dual supertranslation charges \cite{Kol:2019nkc,Godazgar:2018qpq} are akin to the large magnetic charges in QED.
The \emph{global} dual supertranslation charge is a topological invariant on the space of asymptotic metrics, in analogy with the magnetic charge which partitions the space of gauge fields into distinct topological sectors.
The new dual supertranslation symmetry does not correspond to spacetime diffeomorphisms.
Instead, evidence was presented \cite{Kol:2020ucd} suggesting that it is a redundant gauge symmetry of the metric.
In particular, all S-matrix elements are invariant under dual supertranslations and there is no Ward identity associated with them.
As a result, dual supertranslations factorize the Hilbert space of asymptotic states into distinct superselection sectors.
These properties distinguish the new dual supertranslation symmetry from standard supertranslations.

The charges in the \emph{extended BMS group} therefore include supertranslations, dual supertranslations and Lorentz transformations. These charges can be classified according to their transformation law under spatial reflections (parity) into "electric"-types and "magnetic"-types.
The standard and dual supertranslation charge densities are respectively given by the "electric"-type and "magnetic"-type Coulomb components of the gravitational field.
The Coulomb components of the gravitational field are described, in turn, by the leading order terms in the asymptotic expansion of the complex Weyl scalar $\psi_2$.
These leading order monopole charges are analogous to the large electric and magnetic charges in electrodynamics.
In gravity, the dipole moments describe subleading BMS charges and are related to the first subleading terms in the asymptotic expansion of $\psi_2$.
As with the leading order charges, the subleading ones can be classified into "electric" and "magnetic" types according to their parity.
The "magnetic"-type dipole moment describes the Arnowitt, Deser, and Misner (ADM) angular momentum and it generates rotations \cite{ADM,Newman:1968uj,Prior:1977}.
In this paper, we will show that the "electric"-type gravitational dipole moment describes the Beig-\'O Murchadha-Regge-Teitelboim (BORT) center of mass \cite{Beig:1987zz,Regge:1974zd,Chen:2014uma} that generates boosts.
Altogether, the gravitational monopole and dipole moments therefore span the entire extended BMS group.
We describe this structure in table \ref{table:setOfCharges}.

\begin{table}[]
	\centering
	\renewcommand{\arraystretch}{3.5}
	\begin{tabular}{ccc|ccccccc}
		\toprule[1.5pt]
		&\textbf{Multipole Moment} 
		&&&&\textbf{"Electric"-type}&&&  \textbf{"Magnetic"-type}	&
		\\ [2ex] \midrule[1pt]
		&\textbf{Monopole}&&&&\makecell{\text{Supertranslations}\\ \text{(Mass)}}&&&\makecell{\text{Dual Supertranslations} \\ \text{(Dual Mass)}}&
		\\[2ex]\bottomrule[1pt]
		&\textbf{Dipole}&&&&\makecell{\text{Boosts} \\ \text{(Center of Mass)}}&&& \makecell{\text{Rotations} \\ \text{(Angular Momentum)}}&
		\\[2ex]\bottomrule[1pt]
	\end{tabular}
	\vspace{0.4 cm}
	\caption{The extended BMS algebra includes supertranslations, dual supertranslations and Lorentz transformations. This group of symmetries can be classified using two properties: parity and multipole moments. The charge generators of the symmetry operations are mentioned in parentheses.}
	\label{table:setOfCharges}
\end{table}

The dual supertranslation charge gives rise to a duality operation that mixes it with the standard supertranslations generator \cite{Huang:2019cja}.
The structure described in table \ref{table:setOfCharges} therefore suggests that the same gravitational duality operation rotates the dipole generators into each other. In other words, angular momentum and the center of mass are mapped into each other under this operation.
These results suggest that boosts and rotations are related to each other by an "electric-magnetic" type of duality.

The paper is organized as follows.
In section \ref{sec:multipole} we review the multipole expansion of gravitational sources in the Newman-Penrose formalism.
We show how the supertranslation, dual supertranslation and angular momentum charges are related to the multipole moments and discuss concrete examples.
In section \ref{sec:AsyAnalysis} we describe the asymptotic expansion of the metric and of the Einstein equations.
In section \ref{sec:Weyl} we compute the first few leading order coefficients in the asymptotic expansions of the Weyl scalars in terms of the asymptotic data.
Using these results, we then show in section \ref{sec:Lorentz} that the "electric"-type gravitational dipole describes the center of mass, which generates boost transformations on the phase space.
We end in section \ref{sec:Discussion} with a discussion.

\section{Multipole Expansion}\label{sec:multipole}

To set up the stage, we start with a review of the multipole expansion in electrodynamics using the Newman-Penrose formalism.
The electromagnetic case will serve as a familiar example that will be useful to keep in mind when we later study the multipole expansion in gravity.

In the Newman-Penrose formalism \cite{Newman:1961qr,Janis:1965tx,Newman:1966ub,Newman:1968uj}, the electromagnetic field strength $F_{\mu\nu}$ is organized in terms of three complex scalars
\begin{equation}
\begin{aligned}
\phi_0 &= F_{\mu\nu} \ell^{\mu}m^{\nu}, \\
\phi_1 &= \frac{1}{2}F_{\mu\nu} \left(\ell^{\mu}n^{\nu} - \mb^{\mu}m^{\nu}\right),\\
\phi_2 &= F_{\mu\nu} \mb^{\mu}n^{\nu},
\end{aligned}
\end{equation}
where $e^{\mu}_a=\{\ell^{\mu},n^{\mu},m^{\mu},\mb ^{\mu}\}$ is a basis of complex null tetrads in Minkowski background.
The real and imaginary parts of the three scalars encode their transformation law under spatial reflections (parity).
The real parts correspond to field components that are even under parity and therefore of an electric type, while the imaginary parts are odd under parity and therefore are of a magnetic type.

The asymptotic behaviors of the complex Newman-Penrose scalars at large distances are given by
\[
\phi_0 &= \frac{\phi_0^{(0)}}{r^{3}} +\frac{\phi_0^{(0)}}{r^{4}} +\dots ,\\
\phi_1 &= \frac{\phi_1^{(0)}}{r^{2}} +\frac{\phi_1^{(0)}}{r^{3}} +\dots ,\\
\phi_2 &= \frac{\phi_2^{(0)}}{r} +\frac{\phi_2^{(0)}}{r^{2}} +\dots .
\]
Spacetime can then be classified into three zones - near, intermediate and far - according to the different decay rates of the Newman-Penrose scalars.
$\phi_0$ is dominant in the near zone.
$\phi_1$ describes the Coulomb components of the field and is dominant in the intermediate zone.
$\phi_2$ describes the radiative components of the field and is dominant in the far zone.

The physical meaning of the coefficients $\phi_n^{(m)}$ in the expansion of the Weyl scalars was studied in \cite{Janis:1965tx}. The first few leading coefficients correspond to the following solutions:
\begin{enumerate}
	\item Monopole
	\[
	\phi_0 &= 0, \\
	\phi_1 &= \frac{a_0}{r^2},\\
	\phi_2 &= 0.
	\]
	\item Dipole
	\[\label{EMdipole}
	\phi_0 &= \frac{a_1(u)\sin \theta}{r^3}, \\
	\phi_1 &= -\frac{\sqrt{2} \dot{a}_1 \cos \theta}{r^2}  -\frac{\sqrt{2}a_1\cos \theta}{r^3},\\
	\phi_2 &= -\frac{\ddot{a}_1 \sin \theta}{r} - \frac{\dot{a}_1 \sin \theta}{r^2} -\frac{a_1\sin \theta}{2r^3}.
	\]
	\item Quadrupole
	\[
	\phi_0 &= \frac{\dot{a}_2 \sin \theta \cos \theta}{2r^3}
	+ \frac{a_2(u)\sin \theta \cos \theta}{r^4}, \\
	\phi_1 &= 
	- \frac{\ddot{a}_2 (3 \cos^2 \theta -1)}{6\sqrt{2}r^2}
	-\frac{\dot{a}_2 (3 \cos^2 \theta -1)}{2\sqrt{2}r^3}
	-\frac{a_2(3\cos^2 \theta-1)}{2\sqrt{2}r^4}
	,\\
	\phi_2 &= 
	-\frac{a^{(3)}_2 \sin \theta \cos \theta}{6r}
	-\frac{\ddot{a}_2 \sin \theta \cos \theta}{2r^2}
	-\frac{\dot{a}_2 \sin \theta \cos \theta}{4r^3}
	-\frac{a_2\sin \theta \cos \theta}{2r^4}.
	\]
\end{enumerate}
Here $a_0$ is constant while $a_1(u),a_2(u)$ are functions of the retarded null time $u=t-r$. The dot above a letter represents a derivative with respect to retarded null time $\dot{a}_n\equiv\pa_u a_n$.
The real and imaginary parts of $a_0$ are proportional to the electric and magnetic charges, respectively.
In a covariant form these charges are given by the real and imaginary parts of
\begin{equation}\label{EMcharges}
q+ i g = \int_{S^2} d\Omega \, \phi_1^{(0)} ,
\end{equation}
where $q$ and $g$ are the electric and magnetic charges. Large Gauge Transformations (LGT) are then generated by
\begin{equation}\label{LGT}
q(f)+ i g(f)= \int_{S^2} d\Omega \, f(\Omega) \, \phi_1^{(0)} ,
\end{equation}
where $f(\Omega)$ is a real function of the angles on the sphere \cite{Strominger:2013lka,He:2014cra,Campiglia:2015qka,Strominger:2017zoo}.
Similarly, the real and imaginary parts of $a_1$ and $a_2$ correspond to the electric and magnetic components of the dipole and the quadrupole moments, respectively.

Here we would like to make a comment about the interplay between LGT and the multipole expansion of $\phi_1$.
The LGT transformation parameter $f(\Omega)$ can be expanded in multipoles as well.
However, it is important to notice that these two multipole expansions are different.
For example, evaluating the charges in \eqref{LGT} on the dipole solution \eqref{EMdipole} we have
\begin{equation}
q(f_{\ell,m})+ i g(f_{\ell,m})  \sim \dot{a}_1 \delta_{\ell 1}\delta_{m0},
\end{equation}
where $f_{\ell,m}$ are the multipole moments of $f(\Omega)$.
We therefore see that by choosing $f(\Omega)$ with the spherical modes $\ell=1$ and $m=0$, the resulting large gauge charges are proportional to $\dot{a}_1$, which is the time derivative of $\phi_1$'s dipole moment and not the  dipole moment $a_1$ itself.
Therefore we should distinguish the two multipole expansions, though there is a clear interplay between the two.

After reviewing the more familiar case of electrodynamics we will now describe the multipole expansion in gravity using the Newman-Penrose formalism.
The lessons learned from the electromagnetic case will serve as a guidance when analogous concepts are introduced in the gravitational case.
In the Newman-Penrose formalism, the Weyl tensor $C_{\mu\nu\rho\sigma}$ is organized in terms of five complex scalars
\begin{equation}
\begin{aligned}
\psi_0 &= C_{\mu\nu\rho\sigma}
\ell^{\mu}m^{\nu}\ell^{\rho}m^{\sigma}
,\\
\psi_1 &= C_{\mu\nu\rho\sigma}
\ell^{\mu}n^{\nu}\ell^{\rho}m^{\sigma}
,\\
\psi_2 &=C_{\mu\nu\rho\sigma}
\mb^{\mu}n^{\nu}\ell^{\rho}m^{\sigma}
,\\
\psi_3 &= C_{\mu\nu\rho\sigma}
\mb^{\mu}n^{\nu}\ell^{\rho}n^{\sigma}
,\\
\psi_4 &=C_{\mu\nu\rho\sigma}
\mb^{\mu}n^{\nu}\mb^{\rho}n^{\sigma}.
\end{aligned}
\end{equation}
Here $ \{ \ell^\mu,n^{\mu},m^{\mu} , \mb^{\mu} \}$ is a basis of complex null tetrads in the curved background.
As in the electromagnetic case, the real and imaginary parts of the five Weyl scalars encode their transformation law under spatial reflections.
The real parts correspond to field components that are even under parity and therefore of an "electric"-type, while the imaginary parts are odd under parity and therefore are of a "magnetic"-type.

The five complex Weyl scalars describe the ten independent components of the Weyl tensor.
They can be expanded at large distances as
\begin{equation}\label{WeylExp}
\begin{aligned}
\psi_0 &= \frac{\psi_0^{(0)}}{r^5}+ \frac{\psi_0^{(1)}}{r^6} +\dots ,\\
\psi_1 &= \frac{\psi_1^{(0)}}{r^4}+ \frac{\psi_1^{(1)}}{r^5} +\dots , \\
\psi_2 &= \frac{\psi_2^{(0)}}{r^3}+ \frac{\psi_2^{(1)}}{r^4} +\dots , \\
\psi_3 &= \frac{\psi_3^{(0)}}{r^2}+ \frac{\psi_3^{(1)}}{r^3} +\dots , \\
\psi_4 &= \frac{\psi_4^{(0)}}{r}+ \frac{\psi_4^{(1)}}{r^2} +\dots .
\end{aligned}
\end{equation}
One could therefore classify spacetime into five regions according to the decay rates of the Weyl scalars. $\psi_0$ and $\psi_1$ describe near-zone fields and are dominant at small distances. $\psi_2$ describes the Coulomb components of the gravitational field and is dominant in the intermediate zone. $\psi_3$ and $\psi_4$ describe the radiative components of the field and are dominant at large distances.

In addition to the five complex Weyl scalars $\psi_n$, the Newman-Penrose formalism is described by twelve complex spin coefficients which describe the change in the tetrad from point to point.
Of these twelve spin coefficients, the shear
\begin{equation}\label{shear}
\sigma\equiv-m^{\mu}m^{\nu} \nabla_{\nu}\ell_{\mu}
\end{equation}
plays a significant role in the analysis of asymptotically flat spacetimes.
The shear can be expanded at large distances
\begin{equation}\label{asyShear}
\sigma= \frac{\sigmaz}{r^2} + \mO(r^{-3}),
\end{equation}
where $\sigmaz$ is called the \emph{asymptotic shear}. The time derivative of the shear is called the \emph{news} and it characterizes gravitational radiation.

The physical meaning of the coefficients $\psi_n^{(m)}$ in the expansion of the Weyl scalars was studied in \cite{Janis:1965tx}. The few leading coefficients correspond to the following solutions:
\begin{enumerate}
	\item Monopole
	\[
	\psi_0 &= 0, \\
	\psi_1 &= 0, \\
	\psi_2 &= \frac{a_0}{r^3},\\
	\psi_3 &= 0, \\
	\psi_4 &= 0,
	\]
	where $a_0$ is a complex constant.
	\item Dipole
	\[\label{gravDipole}
	\psi_0 &= 0, \\
	\psi_1 &= -\frac{a_1 \sin \theta}{r^4}, \\
	\psi_2 &= \frac{\sqrt{2} \dot{a}_1 \cos \theta}{r^3}
	+\frac{\sqrt{2}a_1 \cos \theta}{r^4}
	\\
	\psi_3 &= 
	\frac{\dot{a}_1 \sin \theta}{r^3}
	+\frac{a_1 \sin \theta}{2r^4}
	, \\
	\psi_4 &= 0, 
	\]
	where $\dot{a}_1$ is a complex constant.
	\item Quadrupole
\[
\psi_0 &= \frac{3a_2 \sin^2 \theta}{r^5}, \\
\psi_1 &= 
-\frac{3\dot{a}_2\sin2\theta}{\sqrt{2} r^4}
-\frac{6 a_2 \sin2 \theta}{\sqrt{2} r^5}, \\
\psi_2 &= 
\frac{\ddot{a}_2(1+3\cos2\theta)}{2r^3}
+\frac{3\dot{a}_2 (1+3\cos2\theta)}{2r^4}
+\frac{3a_2 (1+3\cos2\theta)}{2r^5}
\\
\psi_3 &= 
\frac{ a_2^{(3)}\sin 2\theta}{\sqrt{2} r^2}
+\frac{3\ddot{a}_2\sin 2\theta}{\sqrt{2} r^3}
+\frac{9\dot{a}_2\sin2\theta}{\sqrt{2}2r^4}
+\frac{3a_2\sin2\theta}{\sqrt{2} r^5}
, \\
\psi_4 &= 
\frac{a_2^{(4)}\sin^2\theta}{2r}
+\frac{a_2^{(3)}\sin^2\theta}{r^2}
+\frac{3\ddot{a}_2\sin^2\theta}{2r^3}
+\frac{3\dot{a}_2\sin^2\theta}{2r^4}
+\frac{a_2\sin^2\theta}{4r^5},
\]
where $a_2^{(n)}$ is the $n$-th derivative of $a_2(u)$ with respect to $u$.
\end{enumerate}

The real part of $a_0$ corresponds to an "electric"-type monopole.
An explicit realization of the "electric"-type monopole is the Schwarzschild solution, for which
\[
a_0 &=m, \\
\sigmaz&=0,
\]
where $m$ is the standard Schwarzschild mass parameter.
The imaginary part of $a_0$ is a "magnetic"-type monopole.
An example for a solution with a "magnetic"-type monopole is the Taub-NUT metric, for which
\[\label{TN}
a_0 &= m+i \ell, \\
\sigmaz &=\ell \frac{1+\tan^4\frac{\theta}{2}}{2\tan^2\frac{\theta}{2}},
\]
where $\ell$ is known as the NUT parameter and $\theta$ is the polar angle on the two sphere.
The singularity of the shear at $\theta=0$ and $\theta=\pi$ is due to the Misner string \cite{Kol:2019nkc}.

In covariant form, the monopole charges are given by
\begin{equation}\label{massCharges}
 M+ i \Mt = \frac{1}{4\pi G} \int_{S^2} d\Omega\,  \left(\psi_2^{(0)} + \sigma^0 \pa_u \sigmab^0 \right) ,
\end{equation}
corresponding to the mass and the dual mass, respectively.
Here $G$ is Newton's constant.
$M$ is the ADM mass, which coincides under some assumptions with the Komar mass \cite{Komar:1958wp} and $\Mt$ is the dual Komar mass. For an extensive discussion about the mass and the dual mass we refer the reader to \cite{Kol:2020zth}.
The formula \eqref{massCharges} is akin to the Coulomb charges in electrodynamics \eqref{EMcharges}.
On the Taub-NUT solution \eqref{TN}, the charges \eqref{massCharges} are then evaluated to be 
\[
M &=\frac{m}{G}, \\
\Mt &= \frac{\ell}{G}.
\]
The Taub-NUT solution is therefore characterized by both "electric"-type and "magnetic"-type monopole moments.
The Schwarzschild solution is recovered when the NUT parameter is set to zero.
A pure "magnetic"-type monopole is recovered similarly by setting the Schwarzschild mass $m$ to zero.

Next, $a_1$ represents a gravitational dipole moment, whose real and imaginary parts correspond respectively to "electric"-type dipole and "magnetic"-type dipole.
An example for a solution that contains a "magnetic"-type dipole is the Kerr solution, for which
\[\label{Kerr}
a_0 &=m, \\
a_1 &= i \frac{3G}{\sqrt{2}}  J,\\
\sigmaz&=0,
\]
where $J$ is Kerr's angular momentum
\begin{equation}\label{KerrAM}
J\equiv\frac{a m}{G} .
\end{equation}
Here $a$ is the spin parameter. 
It is therefore evident that the "magnetic"-type gravitational dipole moment describes angular momentum, which is a conserved charge in gravity.
Note that while in electrodynamics only the monopole moments are conserved, in gravity the dipole moments are conserved as well.
From a field theory perspective this could seem strange, but we shall not forget that gravity is a theory of Lorentz transformations.

The generalization of the Kerr angular momentum \eqref{KerrAM} to a covariant expression for the charge is known as the ADM angular momentum. In terms of the Newman-Penrose coefficients it is given by the following integral over $\psi_2^{(1)}$ and the shear
\begin{equation}\label{angularMomentum}
J (\Phi) = -\frac{1}{16\pi G} \int_{S^2} d \Omega \, 
\Phi (\Omega) \, \Im  \left( \psi_2^{(1)} +\eth\left(\sigmazb \bar{\eth}\sigmaz\right)\right).
\end{equation}
The differential operators $\eth,\ethb$ encode the covariant derivative on the sphere and will be discussed soon in detail.
$\Phi$ is an $\ell=1$ spherical mode and the integral in \eqref{angularMomentum} is over a large two sphere.
The first term in \eqref{angularMomentum} was suggested to describe angular momentum in the linearized theory already by Newman and Penrose \cite{Newman:1968uj}. However, to the best of our knowledge Prior \cite{Prior:1977} was the first to derive the exact expression for angular momentum using the Weyl scalars in the non-linear theory.
The angular momentum \eqref{KerrAM} for the Kerr solution \eqref{Kerr} is reproduced with
\[
\Phi(\Omega)= -8 \sqrt{\frac{\pi}{3}}Y_{\ell=1,m=0}(\theta,\varphi)=-4 \cos \theta,
\]
but more generally it can point in any other direction.
In addition to the "magnetic"-type dipole moment, the Kerr solution also contains an "electric"-type monopole moment. In principle, one could isolate the dipole moment by taking the limit $m\rightarrow0$ while keeping the angular momentum $J=am$ fixed.
From the asymptotic point of view it is natural to consider such a limit, but since it violates the bound $a\leq m$, a naked singularity will develop in the bulk.

Finally, an example for a solution that contains an "electric"-type dipole moment is given by the Kerr-Taub-NUT metric \cite{Emond:2020lwi}, for which
\[\label{KTN}
\psi_2 &= \frac{m+i\ell}{(r-i\ell -ia\cos \theta)^3},\\
\sigmaz &=\ell \frac{1+\tan^4\frac{\theta}{2}}{2\tan^2\frac{\theta}{2}}.
\]
The first few leading orders' coefficients in the multipole expansion are therefore given by
\begin{equation}\label{AsyKTN}
\begin{aligned}
\psi_2^{(0)} &=m+i \ell,
\\
\psi_2^{(1)} &=3ia(m+i \ell) \cos\theta,
\end{aligned}
\qquad \Longrightarrow \qquad
\begin{aligned}
a_0 &=m+i \ell, \\
a_1 &= - \frac{3G}{\sqrt{2}}    \left(   K-  i  J\right)  ,
\end{aligned}
\end{equation}
where we have defined the \emph{dual angular momentum} by \cite{Henneaux:2004jw,Bunster:2006rt,Argurio:2009xr}
\begin{equation}\label{dualAngularMomentum}
K \equiv \frac{a \ell}{G}.
\end{equation}
Evidently, the Kerr-Taub-NUT solution contains all four moments: "electric" monopole, "magnetic" monopole, "electric" dipole and "magnetic" dipole.
When $\ell=0$, the solution reduces to the Kerr metric and correspondingly only the "electric"-type monopole and "magnetic"-type dipole survive.
When $a=0$, the solution reduces to the Taub-NUT metric and correspondingly only the monopole moments survive.
When both $\ell=0$ and $a=0$ the solution reduces to the Schwarzschild metric and only the "electric"-type monopole survives.
When both $m=0$ and $a=0$ the solution reduces to the pure-NUT metric and only the "magnetic"-type monopole survives.
Note that the asymptotic shear is proportional to the NUT parameter and is real. It therefore vanishes for the Schwarzschild and Kerr metrics, but is different from zero in the presence of a dual, "magnetic", mass.

The discussion in this section now naturally leads to the following question: what is the physics of the "electric"-type dipole moment? More precisely - does it give rise to a conserved charge and if yes - what is its action on the phase space?
As we will see, the "electric"-type dipole indeed gives rise to a conserved charge known in the literature as the BORT center of mass \cite{Beig:1987zz,Regge:1974zd,Chen:2014uma}, which generates boosts on the phase space.

\subsection{Symmetry Operations}

We have seen that the monopole and dipole moments of the Weyl scalar $\psi_2$ correspond to four conserved charges: mass, dual mass, angular momentum and the center of mass. Each one of these charge acts differently on the phase space.
First, the mass $M$ in \eqref{massCharges} generates time-translations.
When mass density is smeared over the two sphere using a weight function $f(\Omega)$
\begin{equation}\label{superGenerator}
T(f)=- \frac{1}{4\pi G} \int_{S^2} d\Omega\, f(\Omega) \, \Re \left(\psi_2^{(0)} + \sigma^0 \pa_u \sigmab^0 \right),
\end{equation}
it generates spatial translations if $f(\Omega)$ is an $\ell=1$ spherical mode. When $f(\Omega)$ is a spherical mode with  $\ell>1$, the charge \eqref{superGenerator} generates supertranslations, which are analogues to the large gauge transformation generated by \eqref{LGT} \cite{Bondi:1962px,Sachs:1962wk,Strominger:2013jfa,He:2014laa}.

The dual mass $\Mt$ in \eqref{massCharges}, on the other hand, acts trivially on the phase space \cite{Kol:2019nkc}.
However, when smeared over the two sphere 
\begin{equation}\label{dualsuperGenerator}
\Tt(f)
=
-\frac{1}{4\pi G} \int_{S^2} d\Omega \, f(\Omega) \, \Im\left(\psi_2^{(0)} + \sigma^0 \pa_u \sigmab^0 \right),
\end{equation}
it generates non-trivial dual supertranslation transformations, provided that $f(\Omega)$ contains spherical modes with $\ell>1$ \cite{Godazgar:2018qpq,Kol:2019nkc}. The $\ell=0,1$ modes of dual supertranslations are trivial, which means that there are no "magnetic"-type dual transformations of the standard four translations.

As in the electromagnetic case, we would like to emphasize that two different multipole expansion are being discussed here. The first is the multipole expansion of the Weyl scalar $\psi_2$, while the second is of the transformation parameter $f(\Omega)$.
For example, evaluating the charges in \eqref{superGenerator}-\eqref{dualsuperGenerator} on the dipole solution \eqref{gravDipole} we have
\begin{equation}
 T(f_{\ell,m}) +i \Tt(f_{\ell,m}) \sim \dot{a}_1 \delta_{\ell1}\delta_{m0},
\end{equation}
where $f_{\ell,m}$ are the multipole moments of $f(\Omega)$.
We therefore see that by choosing $f(\Omega)$ with the spherical modes $\ell=1$ and $m=0$, the resulting charges are proportional to $\dot{a}_1$, which is the time derivative of $\psi_2$'s dipole moment and not the dipole moment $a_1$ itself.
For example, when angular momentum is conserved $\Im \dot{a}_1=0$ \cite{Janis:1965tx}.
Therefore we should distinguish the two multipole expansions, though there is a clear interplay between the two.

Angular momentum \eqref{angularMomentum} generates rotations over the phase space and similarly the BORT center of mass generates boosts.
We will therefore see that the gravitational dipole charges generate the entire group of Lorentz transformations.
Those can be further generalized to super-Lorentz transformations \cite{Barnich:2009se,	Barnich:2010eb,Barnich:2011ct,Barnich:2011mi,Troessaert:2017jcm,Campiglia:2014yka,Campiglia:2015yka,Campiglia:2015kxa} by promoting their transformation parameters to include $\ell>1$ spherical modes, but we will not discuss this generalization here.
The four leading gravitational moments - monopoles and dipoles - therefore span the entire extended BMS group.
Our goal now is to study the physics of the "electric"-type gravitational dipole.

\section{Asymptotic Analysis}\label{sec:AsyAnalysis}

A general metric can be written in retarded null coordinates using the Bondi gauge \cite{Bondi:1960jsa,Bondi:1962px,Sachs:1962wk}
\begin{equation}\label{BondiMetric}
ds^2 = - V e^{2\beta} du^2 -2 e^{2\beta} du dr +r^2 g_{AB} \Big( d \Theta^A + \frac{1}{2} U^A du \Big)\Big( d \Theta^B + \frac{1}{2} U^B du \Big),
\end{equation}
where $\Theta^A$ with $A=1,2$ are two coordinates that parametrize a two dimensional compact space and
\begin{equation}\label{gaugeCond}
\pa_r \det g_{AB} =0.
\end{equation}
The gauge condition \eqref{gaugeCond}, together with $g_{rr}=g_{rA}=0$, completely fix the local diffeomorphism invariance.
The metric components $g_{AB}$ can be parameterized \emph{\`a la} Sachs \cite{Sachs:1962wk} as
\begin{equation}\label{SachsSphere}
2g_{AB}d \Theta^A d \Theta^B =
(e^{2\alpha} + e^{2\lambda}) d\theta ^2 
+4 \sin \theta \sinh (\alpha-\lambda) d \theta d \varphi
+\sin ^2 \theta (e^{-2\alpha} + e^{-2\lambda}) d \varphi ^2 
\end{equation}
using the standard angle variables.
In this parameterization the gauge condition \eqref{gaugeCond} is satisfied automatically since the determinant of $g_{AB}$ is equal to $\sin^2 \theta$.
Here $V,\beta,\alpha,\lambda,U^A$ are any six functions of all coordinates and they parametrize the most general solution.

The Bondi metric can be expanded systematically around flat space.
In this expansion, the leading order is the Minkowski metric, which is given in retarded null coordinates by
\begin{equation}
ds^2_{\text{Minkowski}} = -du^2 -2dudr +2 r^2 \gammaflat dz d\zb.
\end{equation}
Here we choose to work with the following complex variables
\begin{equation}
\Theta^A=(z,\zb), \qquad 
z = \tan \frac{\theta}{2}  \,  \, e^{i\varphi}
\end{equation}
to parametrize the coordinates on the two sphere. In this system of coordinates, the metric on the unit two sphere is given by
\begin{equation}
\gamma_{AB} = 
\left(
\begin{matrix} 
0 & \gammaflat  \\
\gammaflat & 0 
\end{matrix}
\right),
\end{equation}
where
\begin{equation}
\gammaflat \equiv \frac{2}{(1+z \zb)^2}.
\end{equation}
Note that in complex coordinates
\begin{equation}
\gamma \equiv \det \gamma_{AB} = \gammaflat ^2.
\end{equation}
Using complex coordinates, the metric \eqref{SachsSphere} then takes the form
\begin{equation}\label{metricParametrization}
\begin{aligned}
g_{AB}d \Theta^A d \Theta^B
&=
\left[ \Big(\sinh (2 \alpha)+\sinh (2 \lambda) 
-2 i \sinh (\alpha-\lambda)\Big)
\frac{\gammaflat \zb dz^2}{2z } +\text{c.c.} \right]
\\&
+ \Big(\cosh (2 \alpha)+\cosh (2 \lambda) \Big)  \gammaflat dz d\zb,
\end{aligned}
\end{equation}

The asymptotic behavior determines the fall-off conditions of the metric components, which can be expanded around $r=\infty$ as follows
\begin{equation}
\begin{aligned}
V(u,r,x^A) &= 1 +\frac{V_0(u,x^A)}{r}+\frac{V_1(u,x^A)}{r^2} +\dots,\\
\beta(u,r,x^A) &=\frac{\beta_0(u,x^A)}{r^2}+\frac{\beta_1(u,x^A)}{r^3} +\dots,\\
U_A (u,r,x^A) &= \frac{U_A^{(0)} (u,x^A) }{r^2}+\frac{U_A^{(1)} (u,x^A) }{r^3} +\dots,\\
g_{AB} &= \gamma_{AB} + \frac{C_{AB}}{r} + \gamma_{AB} \frac{C^2}{4r^2}+\dots .
\end{aligned}
\end{equation}
Here we have defined
\begin{equation}
C^2\equiv C_{AB} C^{AB},
\end{equation}
where
\begin{equation}
C^{AB}\equiv\gamma^{AC}\gamma^{BD}C_{CD}.
\end{equation}
We are using a notation where the indices of $C_{AB}$ are raised and lowered using the metric $\gamma_{AB}$, while otherwise the full metric is used to raise and lower indices. This mixed notation could be confusing but we adopt it since it is common in the literature.
The metric in Sachs' parametrization \eqref{metricParametrization} is then expanded as \cite{Godazgar:2018vmm,Godazgar:2018dvh}
\[
\alpha (u,r,\Theta^A) &= \frac{\alpha_0\left(u,\Theta^A\right)}{r}+ \frac{\alpha_2\left(u,\Theta^A\right)}{r^3}+\dots,\\
\lambda (u,r,\Theta^A) &= \frac{\lambda_0\left(u,\Theta^A\right)}{r}+ \frac{\lambda_2\left(u,\Theta^A\right)}{r^3}+\dots ,
\]
where $\alpha_0$ and $\lambda_0$ are related to $C_{AB}$ by
\begin{equation}
C_{AB} = \gammaflat \left(
\begin{matrix} 
(1-i)\frac{\zb}{z}(\alpha_0 +i \lambda_0 )		 & 			0  \\
0 & 		(1+i) \frac{z}{\zb}  (\alpha_0 -i \lambda_0 )
\end{matrix}
\right).
\end{equation}
In particular, $C_{z\zb}=0$ and therefore
\begin{equation}
C^2 = 2C_{zz}C^{zz}.
\end{equation}

\subsection{The Newman-Penrose formalism}

In the Newman-Penrose formalism the metric is expressed using a basis of complex null tetrads
\begin{equation}
e^{\mu}_a = \{ \ell^\mu,n^{\mu},m^{\mu} , \mb^{\mu} \},
\end{equation}
where $\mb^{\mu}$ is the complex conjugate of $m^{\mu}$ and
\begin{equation}
\ell_{\mu}n^{\mu} = -1,
\qquad
m_{\mu}\mb^{\mu} = 1,
\qquad
m_{\mu}\ell^{\mu}=m_{\mu}n^{\mu}=0.
\end{equation}
The inverse metric is given by
\begin{equation}
g^{\mu\nu} = \eta^{ab} e^{\mu}_a e^{\nu}_b
\end{equation}
with
\begin{equation}
\eta^{ab } =\left(
\begin{matrix} 
0 & -1 & 0 & 0 \\
-1 & 0 & 0 & 0 \\
0 & 0 & 0 & 1 \\
0 & 0 & 1 & 0 
\end{matrix}
\right),
\end{equation}
such that
\begin{equation}
g^{\mu\nu} = - \ell^{\mu} n^{\nu}- \ell^{\nu} n^{\mu} + m^{\mu} \mb^{\nu}+ m^{\nu} \mb^{\mu}.
\end{equation}
The Bondi metric \eqref{BondiMetric} can then be described in the Newman-Penrose formalism using the following basis
\begin{equation}\label{NPtetrads}
\begin{aligned}
\ell^{\mu} &= \delta^{\mu}_r ,
\qquad
n^{\mu} =  e^{-2\beta} \Big( \delta^{\mu}_u -   \frac{1}{2}V\, \delta^{\mu}_r 
+U^A \delta_A^{\mu}
\Big),
\\
m^{\mu} &=
\frac{1}{ r }(\cosh\alpha+i \cosh\lambda)\sqrt{\frac{z}{\zb}}\gammaflatt dz
-\frac{1}{ r }(\sinh\alpha+i \sinh\lambda)\sqrt{\frac{\zb}{z}}\gammaflatt d\zb.
\end{aligned}
\end{equation}

Let us remark that the leading order terms in the asymptotic expansion of the tetrad $m^{\mu}$ are given by
\begin{equation}
m^{\mu} =
  \frac{1+i}{2r} \sqrt{   2\gammaflatt    \frac{z}{\zb}   }  \left( dz
  -\frac{1}{r} C^{\zb\zb} d\zb  + \mO\left(r^{-2}\right)   \right) .
\end{equation}

\subsection{The Shear}

Using the Sachs' parametrization \eqref{metricParametrization} one can evaluate the shear \eqref{shear}
\begin{equation}
\sigma=\frac{1+ i \cosh (\alpha-\lambda)}{2} \pa_r \alpha
-\frac{1- i \cosh (\alpha-\lambda)}{2} \pa_r \lambda.
\end{equation}
The asymptotic shear \eqref{asyShear} is then given in complex coordinates by
\begin{equation}\label{leadingShear}
\sigmaz = \frac{i}{2} \frac{z}{\zb} \gammaflatt C_{zz}.
\end{equation}
We see that the tensor $C_{AB}$ is directly related to the shear and therefore it is called the \emph{shear tensor}.
We also have
\begin{equation}
\sigmaz \sigmazb = \frac{1}{8} C^2.
\end{equation}
For example, in the case of the Kerr-Taub-NUT \eqref{KTN} the asymptotic shear in complex coordinates is given by
\begin{equation}
\sigmaz = \ell \frac{1+|z|^4}{2|z|^2} .
\end{equation}
Notice that the asymptotic shear is real and proportional to the NUT parameter only. It therefore vanishes when the solution reduces the the Schwarzschild or Kerr metrics.

\subsection{Spin-weighted derivatives}

Newman and Penrose \cite{Newman:1961qr} introduced the concept of \emph{spin-weight}.
Under rotations in the plane span by $\{m^{\mu},\mb^{\mu} \}$
\begin{equation}
m^{\mu}\rightarrow e^{i \omega} m^{\mu}, \qquad
\omega \in \mathbb{R},
\end{equation}
a quantity $\eta^{(s)}(z,\zb)$ is said to have spin-weight $s$ if it transforms as
\begin{equation}
\eta^{(s)} \rightarrow  e^{i s \omega} \eta^{(s)}.
\end{equation}
Complex conjugation flips the sign of the spin weight.
The covariant derivative on the two surface is then encoded in the operators
\begin{equation}
\begin{aligned}
\eth \eta^{(s)} &= - \frac{1+i}{2} z\, P^{1-s} \paz \left(P^s \eta^{(s)} \right),
\\
\ethb \eta^{(s)} &= - \frac{1-i}{2} \zb \, P^{1+s} \pazb \left(P^{-s} \eta^{(s)} \right),
\end{aligned}
\end{equation}
where
\begin{equation}
P=- \frac{1+z \zb}{|z|}.
\end{equation}
$\eth$ and $\ethb$ raise and lower, respectively, the spin weight by one unit. The commutator of the two operators obeys
\begin{equation}
\left[\ethb,\eth\right]\eta^{(s)} = \frac{s}{2}R_0\eta^{(s)},
\end{equation}
where $R_0$ is the curvature of the celestial space ($R_0=2$ for the two sphere).
We have summarized the spin-weights of different objects in table \ref{table:spinweight}.

 \begin{table}[]
	\centering
	\renewcommand{\arraystretch}{2.5}
	\begin{tabular}{ccc|cccccccccc}
		\toprule[1.5pt]
		&& & & $\eth$	& $\pa_u$&	$\sigmaz$ &	$\psi_4^{(0)}$ &$\psi_3^{(0)}$&$\psi_2^{(0)}$&$\psi_1^{(0)}$&$\psi_0^{(0)}$
		\\ [2ex] \midrule[1pt]
		&$s$&&&	$1$& $0$&$2$ & $-2$ &$-1$&$0$&$1$&$2$ 
		\\[2ex]\bottomrule[1pt]
	\end{tabular}
	\vspace{0.4 cm}
	\caption{Spin-weights.}
	\label{table:spinweight}
\end{table}

\subsection{Einstein Equations}

With the asymptotic expansion described above we can now analyze the Einstein equations perturbatively. 
We will assume that the following components of the stress-energy tensor fall off at large distances as
\begin{equation}
T_{rA}^M = \mO(r^{-3}),
\qquad
T_{rr}^M = \mO(r^{-6}),
\qquad
T_{ur}^M = \mO(r^{-5}).
\end{equation}

With these assumptions on the fall-off conditions for the stress-energy tensor, the $G_{rA}$ components of the Einstein equations at order $r^{-2}$ imply that
\begin{equation}
U_A^{(0)} = - \frac{1}{2}D^B C_{AB}.
\end{equation}
The $G_{rr}$ component of the Einstein equations at orders $r^{-4}$ and $r^{-5}$ imposes
\begin{eqnarray}
\beta_0 &=& -\frac{1}{32}C^2, \\
\beta_1 &=& 0.
\end{eqnarray}
The $G_{ur}$ component of the Einstein equations at order $r^{-4}$ gives
\begin{equation}
V_1 = -\frac{1}{2}D^AU^{(1)}_A
+\frac{3}{32} \left( \Delta -2 \right)C^2 - \frac{1}{2} (D_{A}C^{AB})(D^CC_{CB})
-\frac{1}{8}D^AC^{BC}D_AC_{BC},
\end{equation}
where the Laplacian on the two sphere is $\Delta \equiv D_AD^A$.

\subsection{The Mass Aspect}

The Bondi mass aspect is defined by
\begin{equation}
m_B (u,\Theta^A) \equiv - \frac{1}{2} V_0(u,\Theta^A)
\end{equation}
Using this definition, the constraint equation $G_{uu}$ at order $r^{-2}$ encodes the time development of the mass
\begin{equation}
\pa_u m_B=+\frac{1}{4}D_AD_BN^{AB} - T_{uu},
\end{equation}
where
\begin{equation}
T_{uu} \equiv \frac{1}{8}N_{AB}N^{AB} + 4\pi \lim_{r\rightarrow\infty} r^2 T_{uu}^M
\end{equation}
and
\begin{equation}
N_{AB} \equiv \pa_u C_{AB}
\end{equation}
is the Bondi news.

\subsection{The Angular Momentum Aspect}

Following the notations of Barnich and Troessaert \cite{Barnich:2009se,Barnich:2010eb,Barnich:2011ct,Barnich:2011mi,Troessaert:2017jcm}, the angular momentum aspect $N_A(u,\Theta^A)$ is defined by
\begin{equation}
U^{(1)}_A= -\frac{2}{3} N_A - \frac{1}{6} C_{AB}D_C C^{BC}.
\end{equation}
The constraint equations $G_{uA}$ at order $r^{-2}$ then encodes the time development of the angular momentum aspect
\begin{equation}
\pa_u N_A=
-\frac{1}{4}D^B \left( D_BD^CC_{AC} -D_AD^CC_{BC} \right)
+\pa_A m
-T_{uA},
\end{equation}
where
\begin{equation}
\begin{aligned}
T_{uA} & \equiv
\frac{1}{4}N_{BC} D_A C^{BC}
+\frac{1}{4}D_B \left(
N_{AC}C^{BC}-C_{AC}N^{BC}\right)
-\frac{1}{16} \pa_A \left(C_{BD}N^{BD}\right)
\\
&+8\pi \lim_{r\rightarrow\infty} r^2 T_{uA}^M.
\end{aligned}
\end{equation}

\section{Weyl Scalars}\label{sec:Weyl}

In section \ref{sec:multipole} we have emphasized the role of the Weyl scalar $\psi_2$ in the discussion about conserved charges. In particular, we have demonstrated that the first two leading orders' coefficients in its asymptotic expansion, $\psi_2^{(0)}$ and $\psi_2^{(1)}$, correspond to the monopole and dipole moments.
Our ultimate goal is therefore to evaluate these two coefficients in terms of the asymptotic data described in the previous section \ref{sec:AsyAnalysis}. We would then be able to associate the multipole moments with conserved charges.

\subsection{Leading Order}

We start by evaluating the leading order coefficients in the expansion of the Weyl scalars \eqref{WeylExp}.
First of all, our main interest is in the Coulomb field, which is given at leading order by
\begin{equation}\label{leadingCoulomb}
\psi_2^{(0)} + \sigmaz\pa_u \sigmazb = -m_B - i \mbt.
\end{equation}
Here we have defined the dual mass aspect by
\begin{equation}
\mbt \equiv \frac{1}{4} D^A D^B \Ct_{AB} 
\end{equation}
and the dual shear tensor as
\begin{equation}
\Ct_{AB} \equiv \tensor{\ep}{_{A}^{C}} C_{CB}.
\end{equation}
The covariant, two-dimensional, Levi-Civita tensor is
\begin{equation}
\ep_{AB}=\left(
\begin{matrix} 
0 & +1 \\
-1 & 0 
\end{matrix}
\right) \sqrt{\gamma},\qquad
\ep^{AB}=\left(
\begin{matrix} 
0 & +1 \\
-1 & 0 
\end{matrix}
\right) \frac{1}{\sqrt{\gamma}}
\end{equation}
and in complex coordinates we then have
\begin{equation}
\tensor{\ep}{_{A}^{B}} =\left(
\begin{matrix} 
i & 0 \\
0 & -i
\end{matrix}
\right),
\end{equation}
The dual mass aspect can then be written explicitly as
\begin{equation}
\mbt = - \frac{i}{4} \left(D_z^2 C^{zz}-D_{\zb}^2C^{\zb\zb}\right) =
\frac{1}{2} \Im \left( D_z^2 C^{zz}\right).
\end{equation}
Note that $\mbt$ is real. In terms of the shear we have
\[
2\, i \, \mbt = \eth ^2 \sigmazb - \ethb ^2 \sigmaz.
\]

In addition to the Coulomb field, we also have
\begin{equation}
\begin{aligned}
\psi_3^{(0)} &= \eth \pa_u \sigmazb, \\
\psi_4^{(0)} &= -\pa_u^2 \sigmazb.
\end{aligned}
\end{equation}
The shear is therefore effectively a double integral of the leading radiation field at large distances.
Finally, there is a set of Bianchi identities involving the rest of the leading order coefficients
\[
\pa_u \psi_3^{(0)} &= -\eth \psi_4^{(0)}, \\
\pa_u \psi_2^{(0)} &= -\eth \psi_3^{(0)} +\sigmaz \psi_4^{(0)}, \\
\pa_u \psi_1^{(0)} &= -\eth \psi_2^{(0)} +2\sigmaz \psi_3^{(0)}, \\
\pa_u \psi_0^{(0)} &= -\eth \psi_1^{(0)} +3\sigmaz \psi_2^{(0)}. 
\]

\subsection{Subleading Order}

At the first sub-leading order we will focus on the coefficient $\psi_2^{(1)}$, which we evaluate to be
\begin{equation}
\psi_2^{(1)} =
D^zN_z
-\frac{3}{2}\gammaflatt (D^zC_{zz})(D^{\zb}C_{\zb\zb})
+\frac{1}{4} D^z \left(C_{zz}D_zC^{zz}\right)
+\frac{3}{64} \Delta C^2.
\end{equation}
In terms of the shear the last equation can be written as
\begin{equation}\label{psi21}
\psi_2^{(1)} +\eth\left(\sigmazb \bar{\eth}\sigmaz\right)
+\eth \sigmaz \ethb \sigmazb
+5 \ethb \sigmaz \eth \sigmazb 
-\frac{3}{2} \eth \ethb\left(\sigmaz \sigmazb\right)
+2\sigmaz \sigmazb
=
 D^z N_z +\frac{1}{64}  \Delta C^2  .  
\end{equation}
We have used that
\begin{equation}\label{complexTerm}
\eth\left(\sigmazb \bar{\eth}\sigmaz\right) = \frac{1}{4}D^{\zb}\left(C_{\zb\zb}D_{\zb}C^{\zb\zb}\right)
\end{equation}
and
\begin{equation}\label{realTerms}
\begin{aligned}
\eth \sigmaz \ethb \sigmazb& =  \frac{1}{8}D^B C_{AB} D_C C^{AC}
-\frac{1}{4} D^{A}\left(C_{AB}D_C C^{BC}\right)
+\frac{1}{16} (\Delta - 4 ) C^2, \\
 \ethb \sigmaz \eth \sigmazb & = \frac{1}{8}D^B C_{AB} D_C C^{AC},	\\
 \eth \ethb\left(\sigmaz \sigmazb\right) &=  \ethb \eth\left(\sigmaz \sigmazb\right)
 =\frac{1}{2} \Delta \left( \sigmaz \sigmazb \right)=\frac{1}{16} \Delta  C^2,\\
 \sigmaz \sigmazb & = \frac{1}{8} C^2.
\end{aligned}
\end{equation}
Note that all the expressions in \eqref{realTerms} are real, while \eqref{complexTerm} is a complex function.
We see that the dipole moment $\psi_2^{(1)}$ depends on the angular momentum aspect $N_A$ and on the asymptotic shear $\sigmaz$.

\section{Lorentz Generators}\label{sec:Lorentz}

In the previous section we evaluated the first two leading orders' coefficients of the Weyl scalar $\psi_2$ in terms of the asymptotic data. We are therefore in a good position to study the relation between these coefficients and the asymptotic charges.

Let us start by reviewing the connection between the leading order coefficient $\psi_2^{(0)}$ and the corresponding leading order charges. Using \eqref{leadingCoulomb} we see that the supertranslation charge \eqref{superGenerator}, corresponding to the "electric"-type monopole moment $\Re \psi_2^{(0)}$, can be evaluated in terms of the Bondi mass
\begin{equation}
T(f)= \frac{1}{4\pi G} \int_{S^2} d^2\Theta \sqrt{\gamma} \, f \, m_B.
\end{equation}
This is, of course, a standard result in the literature.
Similarly, the dual supertranslation charge \eqref{dualsuperGenerator}, corresponding to the "magnetic"-type monopole moment $\Im \psi_2^{(0)}$, can be evaluated in terms of the dual mass aspect
\begin{equation}
\Tt(f)=\frac{1}{4\pi G} \int_{S^2} d^2\Theta \sqrt{\gamma} \, f \, \mbt.
\end{equation}
The result is the dual supertranslation charge discussed in \cite{Kol:2019nkc,Godazgar:2018qpq}.

With the results \eqref{psi21} at hand, we can now address the question: what are the charges associated with the dipole moment $\psi_2^{(1)}$? As discussed in section \ref{sec:multipole}, we already know that the charge associated with the "magnetic"-type dipole moment $\Im \psi_2^{(1)}$ is the angular momentum \eqref{angularMomentum}. Angular momentum, in turn, generates rotations on the phase space, which are part of the Lorentz group.
In the notations of Barnich and Troessaert \cite{Barnich:2009se,Barnich:2010eb,Barnich:2011ct,Barnich:2011mi} the Lorentz generators are given by
\begin{equation}
Q(Y)= \frac{1}{32\pi G} \int_{S^2} d^2\Theta \sqrt{\gamma}\left[
Y^A\left(N_A + \frac{1}{32} \pa_A C^2\right)
\right].
\end{equation}
Here the six generators of Lorentz transformations $Y^A$ obey the conformal Killing equation
\begin{equation}\label{cKe}
D_A Y_B + D_B Y_A = \gamma_{AB} \left( D\cdot Y \right),
\end{equation}
where $D\cdot Y \equiv D_A Y^A$.
This can be further generalized to super-Lorentz transformations if $Y^A$ obeys the conformal Killing equation only \emph{locally}, but we will not discuss this generalization here.
Within the solutions to \eqref{cKe}, the three rotation generators are uniquely defined as the divergence-free subgroup
\begin{equation}
\text{Rotations:} \qquad D\cdot Y =0.
\end{equation}
The three boost generators, on the other hand, are uniquely defined as the curl-free subgroup
\begin{equation}
\text{Boosts:} \qquad  \ep^{AB} \pa_A Y_B =0.
\end{equation}
A general Lorentz generator can therefore be decomposed as \cite{Compere:2019gft}
\begin{equation}\label{LorentzDecomposition}
Y^A = \gamma^{AB} \pa_B \Psi  +\ep^{AB} \pa_B \Phi,
\end{equation}
where $\Psi$ and $\Phi$ are both $\ell=1$ spherical modes that respectively parametrize boosts and rotations.

Using the decomposition \eqref{LorentzDecomposition} of the conformal Killing vectors into boost and rotation generators, we can now rewrite the Lorentz charge as
\[
Q(Y)&= -\frac{1}{32\pi G} \int_{S^2} d^2\Theta \sqrt{\gamma} \, 
\left[
+\Psi  \, D^A
\left(N_A + \frac{1}{32} \pa_A C^2\right)
\right. \\ & \left.  
\qquad \qquad \qquad \qquad  \qquad \, \,\,  +i  \,  \Phi \, \ep^{AB}\pa_B \left(N_A + \frac{1}{32} \pa_A C^2\right)
\right].
\]
To derive the last expression we have used integration by parts.
In complex coordinates the Lorentz charge then takes the form
\[
Q(Y)&= -\frac{1}{16\pi G} \int_{S^2} d^2 \Theta \sqrt{\gamma} \, 
\left[
+\Psi  \,
\Re 
\left( D^z N_z + \frac{1}{64}  \Delta C^2\right)
\right. \\ & \left.  
\qquad \qquad \qquad \qquad  \qquad \, \,\, +i  \, \Phi \, \Im\left( D^z N_z + \frac{1}{64}  \Delta C^2\right)
\right]
.
\]

For brevity of the presentation, let us now define
\begin{equation}
\tilde{\psi}_2^{(1)}
\equiv
\psi_2^{(1)} +\eth\left(\sigmazb \bar{\eth}\sigmaz\right)
+\eth \sigmaz \ethb \sigmazb
+5 \ethb \sigmaz \eth \sigmazb 
-\frac{3}{2} \eth \ethb\left(\sigmaz \sigmazb\right)
+2\sigmaz \sigmazb,
\end{equation}
such that equation \eqref{psi21} takes the compact form
\begin{equation}\label{psi21SHORT}
\tilde{\psi}_2^{(1)}
=
 D^z N_z +\frac{1}{64}  \Delta C^2 .
\end{equation}
It is now evident, based on \eqref{psi21SHORT}, that in complex coordinates the Lorentz generators are given by
\[\label{LorentzCharge}
Q(Y)= -\frac{1}{16\pi G} \int_{S^2} d^2 \Theta \sqrt{\gamma}\, 
\left[
\Psi  \,
\Re  \tilde{\psi}_2^{(1)}  
+i  \, \Phi \, \Im  \tilde{\psi}_2^{(1)} 
\right].
\]
We conclude that the real and imaginary parts of $  \tilde{\psi}_2^{(1)} $ respectively generate boosts and rotations. Altogether, $  \tilde{\psi}_2^{(1)} $ generates the entire Lorentz group.

When $\Psi=0$, we recover the result \eqref{angularMomentum} of Prior for angular momentum that is the generator of rotations
\[\label{rotationsGen}
Q(Y) \Big|_{\Psi=0} = i  J(\Phi) = 
-\frac{i}{16\pi G} \int_{S^2} d^2 \Theta \sqrt{\gamma}\, 
 \, \Phi \, \Im  \tilde{\psi}_2^{(1)} 
,
\]
since
\begin{equation}
\Im    \tilde{\psi}_2^{(1)} 
=
\Im  \left( \psi_2^{(1)} +\eth\left(\sigmazb \bar{\eth}\sigmaz\right)\right).
\end{equation}
When $\Phi=0$, on the other hand, the Lorentz charge \eqref{LorentzCharge} reduces to
\[\label{BORT}
Q(Y) \Big|_{\Phi=0} = K(\Psi) \equiv 
-\frac{1}{16\pi G} \int_{S^2} d^2 \Theta \sqrt{\gamma}\, 
\Psi  \,
\Re   \tilde{\psi}_2^{(1)}  
.
\]
Since $K(\Psi)$ generates boosts it therefore coincides with the BORT center of mass. We therefore see that the "electric and "magnetic" type dipole moments of $\psi_2$ are associated with the the real and imaginary parts of the Lorentz charge
\[Q(Y) = K(\Psi) +i J(\Phi),\]
which in turn coincide with the BORT center of mass and the ADM angular momentum, respectively.

As we did for the angular momentum, we would now like to evaluate the center of mass explicitly for a given solution. As an example, we will use the Kerr-Taub-NUT metric \eqref{KTN}-\eqref{AsyKTN} for which we compute
\[
\tilde{\psi}_2^{(1)} =\psi_2^{(1)} 
+\ell^2 \Big(
4\sin^{-2}\theta \left( 3+3\sin^{-2} \theta-4 \sin^{-4}\theta \right)-11
\Big).
\]
For brevity, we are now using the angle variables on the two sphere since the expression above takes a complicated form in complex coordinates.
As can be seen, the contribution to $\tilde{\psi}_2^{(1)} $ due the shear is singular at $\theta=0,\pi$.
This is nothing but a manifestation of the Misner string.
However, when plugged into the expression for the center of mass \eqref{BORT}, the divergent contributions cancel each other. One way to see this is by regulating the integral around the location of the string at $\theta=0$
\[
\int^{\theta=\epsilon} d^2 \Theta \sqrt{\gamma}\, 
\Psi  \,
\Big(
\tilde{\psi}_2^{(1)} -\psi_2^{(1)} 
\Big)
=\delta_{m,0} \times 
6 \ell^2 \sqrt{\frac{3}{\pi }} 
 \Big(
 \log \epsilon 
+\frac{1}{3\epsilon ^4}
-\frac{5}{18 \epsilon ^2}
-\frac{23}{270 }+\dots
\Big)
\]
where the index $m$ refers to $\Psi$'s azimuthal moment and the ellipsis stand for terms that vanish as $\epsilon\rightarrow 0$. It turns out that by regulating the integral around the location of the string singularity at $\theta=\pi$, exactly the same singular terms arise and therefore
\[\lim_{\epsilon\rightarrow 0}\quad
\int^{\theta=\pi-\epsilon} _{\theta=\epsilon} d^2 \Theta \sqrt{\gamma}\, 
\Psi  \,
\Big[
\tilde{\psi}_2^{(1)} -\psi_2^{(1)} 
\Big]_{\text{Kerr-Taub-NUT}}=0.
\]
Hence we find that for the Kerr-Taub-NUT solution, the center of mass is given by
\[
K(\Psi) =
-\frac{1}{16\pi G} \int_{S^2} d^2 \Theta \sqrt{\gamma}\, 
\Psi  \,
\Re   \psi_2^{(1)} ,
\]
effectively casting out the shear from the expression.
Now, for the particular choice
\[
\Psi(\Omega)= 8 \sqrt{\frac{\pi}{3}}Y_{\ell=1,m=0}(\theta,\varphi)=4 \cos \theta,
\]
the BORT center of mass will coincide with the dual angular momentum defined in \eqref{dualAngularMomentum}. More generally, the center of mass can describe boosts in any other direction.

\section{Discussion}\label{sec:Discussion}

In this paper we have studied the extended BMS group, which includes supertranslations, dual supertranslations and Lorentz transformations, using the Newman-Penrose formalism.
The charges in the extended BMS group are classified according to their parity into "electric" and "magnetic" types.
Supertranslations and dual supertranslations are closely related to the monopole moment $\psi_2^{(0)}$ and therefore we refer to them as the \emph{leading BMS charges}. The first corresponds to the "electric"-type monopole $\Re\psi_2^{(0)}$ while the later corresponds to the "magnetic"-type monopole $\Im\psi_2^{(0)}$.
The dipole moment $\psi_2^{(1)}$, in turn, describes Lorentz transformations.
The "magnetic"-type dipole $\Im\psi_2^{(1)}$ corresponds to the angular momentum, which generates rotations over the phase space. Our main result is that the "electric"-type dipole $\Re\psi_2^{(1)}$ corresponds to the BORT center of mass, which generates boosts. Altogether, the gravitational monopole and dipole moments span the entire extended BMS group.

We have studied the Kerr-Taub-NUT metric as an example for a solution that contains all the first two leading multipole moments. In this case, the "electric"-type monopole corresponds to the Schwarzschild ADM mass, the "magnetic"-type monopole corresponds to the NUT parameter, the "magnetic"-type dipole corresponds to the Kerr angular momentum and finally, the "electric"-type dipole correspond to the BORT center of mass (which is sometimes referred to as the dual angular momentum). The charges of the Kerr-Taub-NUT solution are summarized in table \ref{table:KTN}.

\begin{table}[]
	\centering
	\renewcommand{\arraystretch}{2.5}
	\begin{tabular}{ccc|ccccccc}
		\toprule[1.5pt]
		&\textbf{Multipole Moment} 
		&&&&\textbf{"Electric"-type}&&&  \textbf{"Magnetic"-type}	&
		\\ [2ex] \midrule[1pt]
		&\textbf{Monopole}&&&&\makecell{$M= \frac{m}{G}$}&&&\makecell{$\Mt = \frac{\ell}{G},$}&
		\\[2ex]\bottomrule[1pt]
		&\textbf{Dipole}&&&&\makecell{$K=\frac{a\ell}{G}$}&&& \makecell{$ J= \frac{am}{G}$}&
		\\[2ex]\bottomrule[1pt]
	\end{tabular}
	\vspace{0.4 cm}
	\caption{Asymptotic charges of the Kerr-Taub-NUT solution. The leading BMS charges are described by the monopole moments of the Weyl scalar $\psi_2$ while the subleading BMS charges are described by its dipole moment. At each order the charges can be classify into "electric" and "magnetic" types according to their parity.}
	\label{table:KTN}
\end{table}

Let us comment on the relation of our results to the works \cite{Godazgar:2018vmm,Godazgar:2018dvh} (see also \cite{Oliveri:2020xls}), where subleading BMS charges in the Newman-Penrose formalism were studied using the Barnich-Brandt prescription \cite{Barnich:2001jy}.
The first subleading BMS charges, which are associated with the dipole moments $\psi_2^{(1)}$ were dismissed in \cite{Godazgar:2018vmm,Godazgar:2018dvh} as total derivatives.
Indeed, the charge densities of $K(\Psi)$ and $J(\Phi)$ in \eqref{rotationsGen} and \eqref{BORT} are given by total derivative terms and would integrate to zero if the transformation parameters $\Psi$ and $\Phi$ were taken to be constants.
However, when $\Psi$ and $\Phi$ are given by the $\ell=1$ spherical modes, the corresponding charges do not integrate to zero and, as we have shown in this paper, their action on the phase space correspond to Lorentz transformations.
We believe that taking careful consideration of total derivative terms in the Barnich-Brandt prescription will yield the correct Lorentz charges at the first subleading order.

We conclude with some proposals for future research directions.
In General Relativity, the Lorentz generators do not commute with supertranslations, which parametrize in turn a moduli space of vacua. For this reason, the angular momentum of the vacuum is not well defined in General Relativity \cite{Chen:2014uma,Javadinazhed:2018mle}.
Here we would like to suggest that the discovery of dual supertranslations, the resulting "electric-magnetic"-type of duality and the multipole structure of charges could potentially shed new light on this old problem.
This direction is particularly appealing given that angular momentum is described by a \emph{"magnetic"}-type dipole moment.
Perhaps, addressing this question in the extended Einstein-Cartan theory could lead to a better understanding of the issue, in the spirit of the resolution suggested in \cite{Kol:2020zth} for the "magnetic"-type monopole.
Finally, it would be interesting to understand the subleading "electric-magnetic"-type duality in relation to soft theorems (along the lines of \cite{Conde:2016csj,Conde:2016rom}) and to dressing of asymptotic states \cite{Choi:2017bna,Choi:2017ylo,Choi:2018oel,Choi:2019rlz,Choi:2019fuq,Choi:2019sjs}.

\acknowledgments

I would like to thank William Emond, Yu-tin Huang, Reza Javadinezhad, Nathan Moynihan, Donal O'Connell, Massimo Porrati, Jacob Sonnenschein and Shimon Yankielowicz for useful discussions on related subjects.



\bibliographystyle{JHEP}
\bibliography{bibliography}
\end{document}